\documentclass[runningheads,10pt]{llncs}
\usepackage[T1]{fontenc}
\usepackage{graphicx}
\usepackage{todonotes}
\usepackage{hyperref}
\usepackage{paralist}
\usepackage{subcaption}
\usepackage{array}
\usepackage{tabularx}
\usepackage{paralist}
\usepackage{booktabs}
\usepackage{amssymb}
\usepackage{multirow}
\DeclareCaptionLabelFormat{cont}{#1~#2\alph{ContinuedFloat}}
\begin{document}
\title{``I will never pay for this''}
\subtitle{Perceptions of fairness and factors affecting behaviour on `pay-or-ok' models}
%
%
\author{Victor Morel\inst{1}\orcidID{0000-0001-9482-8906} \and
Farzaneh Karegar\inst{2}\orcidID{0000-0003-2823-3837} \and
Cristiana Santos\inst{3}\orcidID{0000-0003-0712-2038}}

\authorrunning{Morel et al.}

\institute{Chalmers University of Technology and University of Gothenburg, Gothenburg, Sweden, \email{morelv@chalmers.se}
\and
Karlstad University, Karlstad, Sweden, \email{farzaneh.karegar@kau.se}
\and
Utrecht University, Utrecht, The Netherland, \email{c.teixeirasantos@uu.nl}}
\maketitle              
\begin{abstract}
The rise of cookie paywalls (`pay-or-ok' models) has prompted growing debates around the right to privacy and data protection, monetisation, and the legitimacy of user consent. Despite their increasing use across sectors, limited research has explored how users perceive these models or what shapes their decisions to either consent to tracking or pay. 
To address this gap, we conducted four focus groups (with \textit{n = 14} participants) to examine users' perceptions of cookie paywalls, their judgments of fairness, and the conditions under which they might consider paying, alongside a legal analysis within the EU data protection legal framework. 

Participants primarily viewed cookie paywalls as profit-driven, with fairness perceptions varying depending on factors such as the presence of a third option beyond consent or payment, transparency of data practices, and the authenticity or exclusivity of the paid content. Participants voiced expectations for greater transparency, meaningful control over data collection, and less coercive alternatives, such as contextual advertising or “reject all” buttons. Although some conditions, including trusted providers, exclusive content, and reasonable pricing, could make participants consider paying, most expressed reluctance or unwillingness to do so.

Crucially, our findings raise concerns about economic exclusion, where privacy and data protection might end up becoming a privilege rather than fundamental rights. Consent given under financial pressure may not meet the standard of being freely given, as required by the GDPR. To address these concerns, we recommend user-centred approaches that enhance transparency, reduce coercion, ensure the value of paid content, and explore inclusive alternatives.
These measures are essential for supporting fairness, meaningful choice, and user autonomy in consent-driven digital environments.


\end{abstract}
\section{Introduction}\label{intro}

The shift from free digital content to paid models has triggered significant debate, particularly concerning the `pay-or-ok' model (also known as \textit{cookie paywalls})\footnote{In this paper, the terms `pay-or-ok' and cookie paywalls are used interchangeably.}. 
‘Pay-or-ok’ models, as conceptualised by the European Data Protection Board (EDPB)~\cite{edpb_opinion_2024}
can be defined as models where a data controller offers data subjects a choice between at least two options in order to gain access to an online service that the controller provides. The data subject can either
1) consent to the processing of their personal data by tracking technologies for a specified purpose, generally, or for behavioural advertising purposes; or 
2) decide to pay a fee and gain access to the online service without their personal data being processed for behavioural advertising purposes. 
 Meta's `pay-or-ok' model implemented in 2023, giving users the choice between consenting to the use of Facebook and Instagram with targeted ads or paying a monthly subscription for an ad-free service~\cite{Meta-payorok}, was recently considered unlawful under the Digital Markets Act (DMA). It was determined that those users who do not consent must have access to a less personalised but equivalent alternative~\cite{EUCOM-Meta}. 
`Pay-or-ok' used to be but are no longer confined to news websites (still amounting to 21.4\% of websites); they have expanded across a broad range of categories, including business (7.7\%), technology (7.3\%),  and are increasingly appearing in areas closer to our everyday lives, such as entertainment (4.6\%), health and medicine (2.7\%), leisure and recreation websites (3.6\%)~\cite{stenwreth2024or}. 

Moreover, the use of this model is becoming prevalent in the EU. Although this model is mostly present in Germany, which hosts 633 of the 805 websites detected by Stenwreth et al.~\cite{stenwreth2024or}, other EU countries such as France (38 detected websites), Spain (33), Italy (29), and Austria (26) were also reported to host such models~\cite{stenwreth2024or}. 

 A growing body of research has examined consent banners,  particularly focusing on their prevalence and compliance~\cite{Sant_etal_20_TechReg,Biel-etal-24-JOLT,ndss19,imc20,acceptall,4years,Matt-etal-20-APF,do_cookie_banners_respect}, user's perceptions~\cite{factorInfluenceConsent-22,Lin-purposes,consentframing-22,dark_patterns_after_gdpr,utz_informed_2019,ndss19,KulykOksana-cookieperception}, and their relation to deceptive designs (also called dark patterns)~\cite{dark_pattern_legal_req,nordichi20,popets2020,asiaccs19,do_cookie_banners_respect,dark_patterns_after_gdpr,taming-cookie-monster,AW-24-popets}.
In contrast, ‘pay-or-ok’ models have received comparatively less attention. 
 Existing studies have primarily addressed their prevalence~\cite{morel2022your,morel2023legitimate,rasaii2023thou} and lawfulness~\cite{morel2022your,morel2023legitimate,Meta-payorok,Bachelet-contractPayorOk,Kollmann-pay-or-ok,edpb_opinion_2024}. 
To the best of our knowledge, only one study reported that 99\% of users consent to targeted ads~\cite{muller2024paying}.
 However, no prior research has explored the decision-making processes users engage in when faced with these binary choices of `pay-or-ok', nor the underlying factors influencing their decisions.

Considering this gap, this paper investigates users' perceptions and expectations of cookie paywalls, and explores whether certain factors might ultimately 
encourage them to pay -- as the payment decision 
could enhance the protection of their personal data and right to privacy, demonstrate the validity of consent, and could financially benefit publishers~\cite{muller2024paying}.
We defined two research questions (RQs) to guide our research: 
\begin{enumerate}
    \item[\textbf{RQ1}] How do users perceive the objectives behind cookie paywalls, the choices they present, and what they expect from them? 
    \item[\textbf{RQ2}] What factors influence users' decisions to either pay or consent when faced with a cookie paywall, and how do these factors impact their decision-making process?
\end{enumerate}
To address our RQs, we conducted a series of four online focus groups 
with a total of \textit{n=14} participants to explore the mental models of users towards cookie paywalls, through a contextualization supported by scenarios and mockups.

Our research reported in this paper makes the following contributions to the literature:
\begin{itemize} 

    \item We provide \textbf{the first in-depth qualitative study of users’ perceptions of cookie paywalls}, offering rich insights into how users understand their purpose, including perceived motivations such as monetisation, legal compliance, or data protection and privacy.
    \item We reveal how \textbf{users' perceptions of fairness are highly conditional on exogenous and hypothetical scenarios} -- they deemed cookie paywalls fair if transparent, with authentic content, and considering they could exit that website -- but often leading to conflicting but deeply reasoned judgments.
    \item We identify \textbf{key decision-making factors} potentially influencing whether users choose to pay, consent, or exit the websites exposing them to cookie paywalls, including exclusive content, pricing, and trust in providers; although ultimately, no killer feature would lead users to pay. 
    \item We contribute \textbf{user-centred and policy recommendations}, including calls for greater transparency, the inclusion of meaningful alternatives (e.g., contextual ads and a ``reject all'' option), and mechanisms to ensure content authenticity, while raising critical questions about the ethics and lawfulness of consent in economically constrained contexts, and ultimately, about the legitimacy of this model.
    \end{itemize}

By exploring these dimensions, our study offers practical, legal insights for platform designers, regulators, and policymakers seeking to align digital consent mechanisms with user expectations, legal requirements, and fairness in practice.

\section{Methodology}\label{method}

We addressed our research questions through four online, 90-minute focus groups run on Zoom between November 2024 and January 2025, plus one pilot session. 
Focus groups are a widely used method in Human-Computer Interactions relying on qualitative analysis with few participants~\cite{lazar2017research}.
They can provide deeper insights than quantitative methods (such as surveys), while being less labour-intensive than individual interviews.
They also present ``a broad range of viewpoints and insights'', and ``can help overcome many of the
shortcomings of interviews'' such as having non-talkative interviewees~\cite{lazar2017research}.
For the main focus groups, 14 adults (FG1 = 5, FG2 = 4, FG3 = 3, FG4 = 2\footnote{One focus group had only two participants because three registered participants did not show up or notify us in advance, making it impossible to reschedule. Despite the small size, the session still produced valuable insights through active discussion.}) were recruited on Prolific---an online platform for academic research that offers reliable, pre-screened participants and advanced demographic filters---under the following criteria: English fluency, residence in the EU/EEA or Switzerland, and a working microphone; no prior privacy expertise was required. We did not collect participants’ exact countries of residence, in line with the principle of data minimisation, as this information was not necessary for our analysis. Instead, we used Prolific’s screening filters to ensure that all participants were located in the EU/EEA or Switzerland. This approach made it possible to recruit a more diverse sample in terms of both region and age compared to on-campus recruitment, which would have been slower and less diverse. Our setup for Prolific filters also yielded a gender-balanced sample (7 women, 7 men). After consent, participants completed the demographic questionnaire listed in Appendix \ref{demog}.

Our sample size was modest, and we do not claim to have achieved \textit{full theoretical saturation}, the point where no new themes or insights emerge from additional data collection. However, given our focused research aim and consistent inclusion criteria, we approached \textit{code saturation}. The rich 90‑minute discussions and iterative team-based analysis provided meaningful exploratory insights. 

Table~\ref{tab:focusgroup} presents an overview of demographic information. Most participants were young (10/14 were 18–34 years), well-educated (11 held $\geq$ bachelor’s degrees), and came from varied fields, e.g. economics, marketing, early-childhood care, and computer science. Nine were (self-)employed, three were job-seeking, and one was a student; half reported earning < € 1335 per month after tax and 43\% earned € 1335–2225 (one non-response).~\footnote{Currencies have been converted to euros in the paper for consistency.} 
Note that the annual median net income is € 21588 in the EU.~\footnote{\scriptsize \url{https://ec.europa.eu/eurostat/databrowser/view/ilc_di03/default/table?lang=en}}
We cannot provide more details on their economic situations (e.g., the country, cross-checking with age, etc.) due to anonymity reasons and to follow the data minimisation principle.
While all felt confident using computers, self-reported cookie-banner understanding was lower: one participant managed banners effortlessly, seven encountered minor issues, five needed occasional clarification, and one struggled frequently.

\renewcommand{\arraystretch}{1.1} 
\setlength{\tabcolsep}{3pt} 
\begin{table}[ht!]
    \centering
    \small
    \caption{Demographic overview of focus group participants (F: Female, M: Male, HS: High School)}
    \begin{tabularx}{1\linewidth}{|c|l|l|X|}
        \hline
        \textbf{FG} & \textbf{Age Range} & \textbf{Gender} & \textbf{Degree} \\
        \hline
        1 & 18-24: 2, 25-34: 1, 35-44: 1  & F: 1, M: 3 & BSc: 2, HS: 2  \\
        2 & 25-34: 3, 18-24: 1, 35-44: 1  & F: 2, M: 3  & MSc: 1, BS: 3, HS: 1\\
        3 & 25-34: 2, 65+: 1  & F: 2, M: 1  & MSc: 2, BSc: 1  \\
        4 & 35-44: 1, 25-34: 1  & F: 2  & MSc: 1, BSc: 1  \\
        \hline
        \textbf{Total} & 18-24: 3, 25-34: 7, 35-44: 3, 65+: 1  & M: 7, F: 7  & MSc: 4, BSc: 7, HS: 3  \\
        \hline
    \end{tabularx}
    \label{tab:focusgroup}
\end{table}

Sessions followed Krueger’s guidance~\cite{krueger_2002_fginterviews}. Zoom audio was recorded locally and transcribed with Amberscript; transcripts were analysed thematically following Braun \& Clarke’s method \cite{braun_2006_method} (details of our procedure are in Appendix~\ref{theme}). 

\noindent\textbf{Ethical considerations:} The study was approved by Karlstad University’s ethical advisor (HS 2024/1242) and Chalmers University’s Data Protection Officer. Data were handled in compliance with the GDPR. Participants used pseudonyms and disabled cameras during recording. All attendees, including the pilot group, were compensated. Prolific participants were compensated at €10 (£8.67) per hour (€15 / £13.00 for 90 minutes), following Prolific’s recommended rate.

\subsection{Course of the focus group}\label{plan}
Each session followed a four-steps script: (1) an introductory segment that re-explained the focus-group format, confirmed oral consent, and set ground rules; (2) a short briefing on cookie banners and cookie paywalls; (3) an open discussion to gauge participants’ general perceptions and expectations of cookie paywalls; and (4) scenario-based discussions using mock-ups. Participants were thanked and compensated immediately after the session.

\textbf{Presentation of the context and general discussion.}
We began by reminding participants that cookie banners are legally mandated (e.g., GDPR) and distinguishing personalised from contextual advertising. Next, we defined \emph{cookie paywalls}, showed real-world screenshots, and situated them within the broader `pay-or-ok' model, contrasting Meta’s tracking-permissive subscription (the pay option only states that personal data will not be used for ads) with a true cookie paywall that lets users avoid all tracking.
Participants then described (i) their usual reactions to ordinary banners, (ii) any experience with cookie paywalls (did they pay, accept all cookies, or leave?), and (iii) their views on the \emph{purpose} and \emph{fairness} of paywalls compared with regular banners. The discussion was open-ended, but prepared prompts kept it on track when needed.

\textbf{Scenarios and mockups.}
To probe what shapes willingness to pay, we showed six Contentpass-style  -- a leading Subscription Management Platform (SMP) -- mock-ups and asked participants to act as though they had just landed on the page, choosing to either pay, accept tracking, or leave, and to justify that choice. Mockups for all scenarios are presented in Figures~\ref{fig:scenario_0} to~\ref{fig:scenario_5} in Appendix~\ref{appendix:figures}.

\textbf{Scenario 1 – Tracking \& ad-free vs. ad-free only.} Baseline design (SEK 30/€ 2.67 per month). At the first step, we compared a full tracking- \& ad-free option with an ad-free-only option.

\textbf{Scenario 2 – One-click payment.} Same price/features as the baseline, but payment was framed as a single, equally effortless click. We highlighted that in both options (baseline and one-click), technicalities and trust issues were not critical (i.e., payment was safe and working).

\textbf{Scenario 3 – Exclusive content.} Paying also unlocked unique material; participants were encouraged to think of what “exclusive” content would tempt them.

\textbf{Scenario 4 – Cheap price.} The fee dropped to SEK 5/€ 0.45 per month, plus a non-discounted yearly fee (SEK 60 SEK/€ 5.35), so they could weigh different price points.~\footnote{We actively encouraged participants to picture what `their' cheap price was.}

\textbf{Scenario 5 – Transparency on third-party sharing.} Banner explicitly stated data would be sold to “300 + vendors”, whereas the pay option promised full privacy. Participants were asked to picture a site they already trust, so trust itself would not dominate responses.

\textbf{Scenario 6 – Various website types.} Four imagined sites, recipes, sports news, health advice, and e-commerce tested whether data sensitivity or personal interest changes willingness to pay.

For each scenario, we captured the decision and the stated reasoning, providing rich material for the subsequent thematic analysis (see Appendix~\ref{theme}).

\section{Results: user perceptions and decision-making of cookie paywalls}

Our thematic analysis identified four themes (\textbf{TX}: \textbf{T1} to \textbf{T4}), with three addressing RQ1 (reported in Sections~\ref{results:RQ1_1}-\ref{results:RQ1_3}), and one addressing RQ2 (reported in Section~\ref{results:RQ2}).
An abridged version of the codebook (only presenting the most important codes discussed in this section) can be found in Table~\ref{tab:codebook_small}, while the comprehensive codebook can be found in Table~\ref{tab:codebook} in the Appendix.
Codes are labelled as \textbf{TX.X} (e.g., \textbf{T1.2} represents the second code under the first theme). 
Some codes emerged from discussions within specific subsets of focus groups, indicated in the format \textbf{TX.X} [FGX] (e.g., \textbf{T1.2} [FG1, FG2], where FG stands for Focus Group), while others were present across all focus groups, in which case the focus group reference was omitted.

We report verbatim comments from participants (PX), comments have been edited for clarity without altering their meaning.

\begin{table}[!ht]
\caption{Abridged version of the codebook resulting from our thematic analysis.}
\label{tab:codebook_small}
\footnotesize

\begin{tabular}{@{}p{4cm}|p{7.9cm}@{}}
\toprule
\textit{\textbf{Themes}} & \textbf{Codes} \\ \midrule
\multirow{3}{=}{\textbf{\textit{T1}} Users' perceptions of cookie paywalls' objective} & \textbf{T1.1} Monetary purpose for the websites - make money either way \\
 & \textbf{T1.2} To maintain users' privacy \\
 & \textbf{T1.3} To comply with regulations \\ \midrule
\multirow{6}{=}{\textbf{\textit{T2}} Perception of fairness of cookie paywalls} & \textbf{T2.1} Fair as users are not forced to take any options \\
 & \textbf{T2.2} Fair as nothing comes for free \\
 & \textbf{T2.3} Fair if transparent \\
 & \textbf{T2.4} Fair if authentic and original content \\ 
 & \textbf{T2.5} Unfair because forced to make a choice (can’t reject) \\
 & \textbf{T2.6} Unfair as privacy should not be a privilege \\ \midrule
\multirow{4}{=}{\textbf{\textit{T3}} Expectations of cookie paywalls design} & \textbf{T3.1} A third option - Reject all \\
 & \textbf{T3.2} Better ex-ante transparency and control of data practices \\
 & \textbf{T3.3} Additional choice to have contextualized ads only \\
 & \textbf{T3.4} Cookie paywalls should not exist \\ \midrule
\multirow{8}{=}{\textbf{\textit{T4}} Factors affecting behaviour/decisions on cookie paywalls} & \textbf{T4.1} Access to exclusive content/service (regularly) \\
 & \textbf{T4.2} Not paying regardless of the situation \\
 & \textbf{T4.3} Cheap/fair subscription price \\
 & \textbf{T4.4} Not bothered/influenced by ads \\
 & \textbf{T4.5} Get rid of (blocking) ads \\
 & \textbf{T4.6} Feeling of pervasive tracking and data sharing \\
 & \textbf{T4.7} Perceived manipulation of content/design \\
 & \textbf{T4.8} Trust in the service provider \\ \bottomrule
\end{tabular}
\end{table}

\subsection{Perceptions of cookie paywalls objectives}\label{results:RQ1_1}

The first theme related to RQ1 is our participants' \textbf{\textit{perceptions of cookie paywalls' objectives (T1)}}. 
This theme shows participants' perception of service providers' objectives for presenting users with cookie paywalls. The most commonly perceived objective of cookie paywalls was \textbf{monetary purposes for the websites - make money either way (T1.1)}. 
The majority of participants across all focus groups (13 out of 14) believed that, whether users choose to pay or accept cookies, the website ultimately benefits financially. 
As P9 (FG2) explained: \textit{their goal is always the same. It's to make money. It's to make profit [...]. So either they take your data and use it to make their advertising or databases more accurate and more efficient, or they get some money from you [...], a direct monetary compensation}. 
P4 (FG1) also believed that consumers should not be concerned with how service providers generate revenue and saw cookie paywalls as \textit{just additional money on top}.

Nonetheless, participants had differing opinions on whether accepting cookies or paying would be more beneficial for service providers. For instance, while P11 (FG3) believed that accepting cookies is more beneficial for them: \textit{they make more of that [...] they can create a pattern [...] They can see what you're interested in [...] by accepting the cookies, you give them way more like valuable information because they kind of see they can profile you}, P1 (FG1) had a contrasting view, stating \textit{the pay option is a way for the websites to generate a higher average revenue per user because I think they make more money if you pay. But I also know that they know that most people are not going to pay. So they just put the accept all [...] we are going to get targeted ads and they might make commissions}. 



\textbf{To comply with regulations (T1.2 [FG1, FG3])} was another objective mentioned by three participants; this topic came up less often than monetary purposes. P4 (FG1) believed that \textit{they need to give another option besides accepting the cookies legally speaking. And this may be a way to skirt that responsibility or give another option.} P11 (FG3) phrased it as \textit{the illusion that we have a choice or something.}

To our surprise, \textbf{maintaining users' privacy (T1.3 [FG1])} was not an objective that several participants associated with cookie paywalls, and it was only briefly discussed in one focus group. As P1 (FG1) expressed, \textit{you maintain privacy. I think it's about privacy and protecting your data. And you're kind of paying to protect your privacy. And privacy should be a right}.
The pay option was also perceived as a way to meet the needs of certain groups of people who care about their privacy, as P2 (FG2) put it: \textit{[...] some people really care about privacy. So by giving them an option, maybe there are some users that they will get to pay that otherwise would just probably leave the site.}

\subsection{Fairness of cookie paywalls}\label{results:RQ1_2}


Another area of discussion, included in the focus group script, was participants’ \textbf{\textit{perception of the fairness of cookie paywalls.}} Participants had varied and opposing opinions about the fairness of cookie paywalls. Paywalls were perceived as fair for different reasons and under different conditions. Interestingly, some participants, P1 and P2 (FG1), and P6 and P8 (FG2), even expressed contradictory views, perceiving the paywalls as both fair and unfair depending on contextual factors such as the availability of alternatives or clarity of information. Among the 14 participants, 10 referred to aspects that made cookie paywalls seem unfair, and 8 mentioned reasons for why they found them fair. These counts include overlaps; several participants expressed both perspectives, while others only mentioned fairness or only unfairness.

Offering authentic and original content was the most frequently mentioned justification for the fairness of paywalls (\textbf{fair if authentic and original content (T2.4 [FG2, FG3, FG4])}).
Participants believed that if the content is original, it cannot be found elsewhere, it is fair to monetize it, as P9 (FG2) put it: \textit{If it's something they had to put effort and work into, then, of course, it's fair that they should have some compensation for their work.} P8 (FG2) similarly stated that if the content is \textit{revised or there's a team behind it and they have certain quality,} it is fair to expect compensation for it.

Some participants believed that no service or (original) content should be free, as companies offering them have costs to cover, making paywalls fair (\textbf{fair as nothing comes for free (T2.2 [FG1])}). P1 (FG1) and P3 (FG1) framed it respectively: \textit{I know that these are companies and these are people's jobs. And of course, they should be compensated and paid for it,} and \textit{I don't think things should be free. [...] it's people's jobs. They've worked to create that content.} 

Participants believed that paywalls are fair if the options provided are presented honestly and transparently (\textbf{fair if transparent (T2.3 [FG1, FG2])}), as P8 (FG2) put it, \textit{they can really understand what they are agreeing to.}

Interestingly, participants had opposing opinions about whether cookie paywalls genuinely offer a choice. 
Those who believed people are not forced to take any of the options, such as P2 (FG1), who said, \textit{I think they are entitled to put the options in front of us and we either accept or reject [the offer],} deemed paywalls as fair (\textbf{fair as users are not forced to take any options (T2.1 [FG1, FG2, FG3])}).
As P7 (FG2) framed it, \textit{it's a business model, when the company does not receive your cookie, you either pay either leave the website [...] I think it's fair.}
However, those who believed that people do not truly have a choice and are forced to select one of the two options, without the ability to reject all cookies or at least accept only certain cookies based on their preferences, perceived paywalls as \textbf{unfair because forced to make a choice (T2.5)}. As P11 (FG3) put it: \textit{I don't think having to pay or just accept all is fair [...] Everything feels forced. We don't really have a choice}. The perception of a lack of choice, particularly the absence of a “reject all” option to access services or content, was the most frequently mentioned factor in justifying their unfairness.

The requirement to either ``accept all'' or ``pay'' was also associated with being forced to give up privacy, as P14 (FG4) framed it: \textit{pay or you don't have your data privacy and I don't like at all}. Additionally, some participants linked this practice to ``coercion'', as P13 (FG4) explained: \textit{[...] it forces people to actively decide between privacy at a cost and free access with tracking [...] it could also feel like coercion [...]}. Similarly, it was compared to ``blackmail'', particularly when users needed access to essential content or services. As P11 (FG3) described: \textit{They know we're interested [in the information behind the paywall] and we're more likely to accept these absurd conditions. Of course, we're not forced. We can just leave the website, but if we're there, because we probably have to get that information, whatever information it is. So I feel it's a bit like here's a bad option, here's a worse option, you know? So it's almost like  blackmailing.}

Additionally, paywalls were also perceived as \textbf{unfair as privacy should not be a privilege (T2.6 [FG4])}. In FG4, privacy as a fundamental right was discussed in relation to fairness. Participants believed that privacy should not be treated as a privilege that people must purchase separately. As P14 (FG4) noted, \textit{it feels a bit like a trap [...] why should I have to pay to keep my data private? It kind of makes me think that privacy is being turned into a privilege which is super unfair for me}, especially because, as P13 (FG4) emphasised, \textit{paying for privacy is not a viable option for everyone.}

\subsection{Expectations of cookie paywalls}\label{results:RQ1_3}

Our analysis of the focus groups' data revealed another theme related to \textbf{\textit{expectations of cookie paywalls (T3)}}, as referred below.


 The most frequently mentioned expectations across all focus groups were 
 the first two. 
 %
 Participants expressed frustration about being forced to choose between options they did not want and \textbf{wished for the existence of the option to reject all cookies (T3.1)}. As P2 (FG1) and P6 (FG2) put it respectively: \textit{[...] I think all cookie banners should have a reject-all option present. They shouldn't be able to lock you down by checking all the boxes}, and \textit{I feel like I'm forced to pay or to accept, in this cookie paywall. I think there should be an option to reject also the cookie}.


Participants expected clearer and more trustworthy information regarding the options presented to them, along with greater control over what they were required to share (\textbf{better ex-ante transparency and control of data practices (T3.2)}). Particularly, they wanted to understand exactly what data was being tracked, who would have access to it, and what they would gain if they chose to pay. 
Additionally, they questioned why service providers required users to accept ``all cookies'' and why they were not given more control over selecting which cookies to accept. For example, P9 (FG2) believed that \textit{if I have some more information about what's being tracked and what's being taken as data} and \textit{if I knew exactly what was being taken and what was being shared and with who}, they could make more informed decisions.

Participants also emphasised the need for information to be presented in a \textbf{trustworthy} manner that did not feel manipulative or aimed at nudging them toward a specific choice. As P1 (FG1) expressed, \textit{
that wording about your data will be sold to 300 plus vendors seemed a little bit manipulative and, kind of like a threat or blackmail. So I think wording it in an honest way without making it seem like a threat, your data will be sold, is really important [...]}. 

The request for ex-ante transparency was not limited to the ``accept'' option but also applied to the ``pay'' option. Participants expected mechanisms to ensure the exclusivity of content. As further discussed in Section~\ref{results:RQ2}, accessing exclusive content/services was one of the strongest motivations for paying when faced with cookie paywalls. However, ensuring that the content was truly exclusive was crucial, as P9 (FG2) stated: \textit{explain or try to explain better what the exclusive content is or what you have to offer, how it's different from other websites and other sources [...]}. 

In total, six participants from all the focus groups expected an \textbf{additional choice to have contextualised ads only (T3.3)}. 
They believed that contextualised ads could provide a middle ground between consenting to personalised ads and paying not to be tracked, allowing service providers to monetise while respecting users' privacy. 
P8 (FG2) also referred to a business model where users \textit{don't have access to all the content and you get contextualized ads.}, and P14 (FG4) mentioned: \textit{[...] So the website would still be monetised, but I wouldn't have to pay neither with my information or with my money.}

Some participants took a step further, believing that \textbf{cookie paywalls should not exist (T3.4 [FG2, FG3, FG4])}. They criticised the business model and questioned the necessity of paywalls, arguing that \textit{what is happening is not the way they want to engage with the internet [...] It just feels like it's really taking advantage of people.} (P10 (FG3)). One participant (P5 (FG2)) also pointed out that \textit{there are sites that are running and that are earning money and don't have the cookie paywall. So it's possible to run a site without it}.

\subsection{Factors influencing users' pay-or-ok decisions in cookie paywalls}\label{results:RQ2}
Participants mentioned numerous, sometimes contradictory, \textit{\textbf{factors affecting their behaviour/decisions on cookie paywalls}}, a theme (\textit{\textbf{T4}}) addressing RQ2.
Those factors were said by participants to lead them to either 1) accept all cookies, 2) pay, or 3) leave the website if they deemed the choice unacceptable; although some factors were discussed as leading towards \textit{not paying}, without specifying whether it would mean accepting the cookies or leaving the website.
Some factors were impacting participants' decision-making ambivalently: the same factor could be a reason to pay for one participant, and to accept for another.
This section only introduces  \textit{the most cited factors in our focus groups}, cited across three or four focus groups for a total of at least 10 times.

The most prevalent factor was the \textbf{access to exclusive content/service (T4.1)}, heavily mentioned in all focus groups.
Most participants valued the possibility to access otherwise gated content, which would interest them, or that they would need (e.g., for educational or professional purposes).
P9 (FG2) explains \textit{If it's something very exclusive and I can only see in this website, I will probably pay}, and P6 (FG2) \textit{maybe some information that I will have the access to read it earlier than the others users that they don't pay. So yes, in this scenario also, I will pay}.
This factor is a typical example of an ambivalent factor.
For instance, for P2 (FG1), the need to access a specific content is first a reason to accept cookies, but ultimately to access the content: \textit{If it's a website that I really don't need to use, I wouldn't use it. And then if I actually needed to use it first, I guess I would accept the cookies and only as a last resort I would pay.}
But the decision is often pondered, and some participants, such as P8 (FG2), would first want guarantees on the exclusiveness of content: \textit{if I were to pay, I would like to see what that exclusive content entails, if it's really that it's so safe and interesting.}
However, it is important to highlight that in several cases, the exclusive content was the only reason to pay invoked by participants, such as P11 (FG3) \textit{But like I said, if it's something I really need. So that's the only reason why I would still pay.} 

The second most prevalent factor was less a factor impacting the decision-making than a blanket refusal to pay for cookie paywalls.
Indeed, participants defended across all focus groups that they would \textbf{not pay regardless of the situation (T4.2)}.
P10 (FG3) resented the model generally speaking \textit{It's like almost like a moral principle or something. I feel, I couldn't on my own. Yeah. Principles, I know I couldn't pay.}
When facing various mockups introduced in our scenarios, participants consistently replied \textit{Same I wouldn't, I wouldn't pay, I just accept all.} (P3, FG1, regarding the different types of websites).

However, some factors would get participants more inclined to pay, such as a \textbf{cheap/fair subscription price (T4.3)}, a factor often mentioned in combination with other factors.
P13 (FG4) details their conception of a fair price as \textit{I would be willing to pay it if the price feels fair for the value.} 
For P9 (FG2), the price matters, but the content behind the cookie paywall remains crucial: \textit{The price totally makes a difference, of course. But I keep what I said previously. It's all about what the content is.}
Whereas for P2 (FG1), the price is decisive, but so is the removal of ads (we had given a rough estimation in euros): \textit{for less than €5 a year. It would be nice to not have to see all those ads.}
Note that some participants were explicitly relating their choice to their economic situation, such as P12 (FG3) who explains that they would not pay \textit{Well, because I'm very poor}.~\footnote{Recall from the demographics that half of our participants earned an annual net income of less than €16500 a year, for a median net income of €21588 in the EU.}

Ads were seen as an ambivalent factor: towards consent when participants were \textbf{not bothered/influenced by ads (T4.4)}, or towards paying when they wanted to \textbf{get rid of (blocking) ads (T4.5)}.
P1 (FG1) explains: \textit{normally I accept all still because I'm not really a victim of advertising. I'm not prone to buying things, but I actually like seeing ads that are dedicated to me or personalised for me. I don't mind that}, although they add \textit{I do agree that it's a concern for some people.}
P8 (FG2) feels differently and specifies \textit{I don't really like ads, especially if they are intrusive. If I'm reading news or I'm reading a blog or whatever, I don't really like watching any kind of ads.}, a feeling even more prominent in a context where ads cannot be blocked, such as YouTube, which could justify paying.

A \textbf{feeling of pervasive tracking and data sharing (T4.6)} veered participants towards giving consent. 
This feeling made them feel disempowered, and according to them, made illusory the protection of their data and their privacy.
%
P12 (FG3) had given up on not being tracked: \textit{I would accept because ads and tracking are not such an inconvenience, I am not going to pay 3 or €4 a month just to avoid ads and the tracking, I find it almost impossible. I will be tracked anyway.}
According to P7 (FG2), paying for privacy on one website cannot match the data collection of big players: \textit{Well, by my opinion, if you are using the Instagram or the Facebook or the Snapchat or any other messenger, I would say you already are giving your personal data.} 

Scenario 5, which investigated a (hypothetical) website transparent about its data practices, 
backfired and triggered reactance from participants (although the design reflected current practices~\cite{morel2023legitimate}).
Participants experienced a feeling of \textbf{perceived manipulation of content/design (T4.7)}.
It led some participants to believe that the website may want to trick them into paying by overplaying the risks associated with accepting cookies, and eventually to leave the website as a result.
P1 (FG1) expressed that: \textit{Personally, I think the ``your data will be shared with 300 plus vendors'' I know it's being transparent, but I think it's kind of provocative in a way. They're being transparent and they're being honest. But I think it's written in a way that's going to make people want to pay more. I would personally not visit this website.}

Another important factor that inspired participants not to pay is the \textbf{trust in the service provider (T4.8)}.
On the one hand, not enough trust in the payment mechanism inspired aversion to it. 
P3 (FG1) explains: \textit{I more than likely would not pay. It's when you're adding in your card details or giving away your financial information. I have to feel 100\% confident.}. 
%
%
On the other hand, trust could lead to accepting cookies, a possible explanation being that they feel comfortable with data collection which they presumably deem acceptable and non-abusive.
P2 (FG1) shares that: \textit{if the website is not trustworthy and I really don't have to access that information, surely I will not accept the cookies and I will probably not even use the website. But if it's something that I really need to see and I actually kind of trust the website. So most likely I will accept the cookies.}
%
From these perceptions, we note that \textbf{trust never affected participants in a way that would lead them to pay}.

\section{Discussion} 

This section discusses our findings concerning the research questions (Sections~\ref{discussion:rq1} and~\ref{discussion:rq2}), contextualising participants’ perceptions, expectations, and decision-making regarding cookie paywalls. 
We highlight how these perceptions often diverge from the legal and normative assumptions underlying current implementations, especially regarding consent and user autonomy. We conclude by offering design and policy recommendations (Section~\ref{subsec:reco}) to address these gaps and improve the legitimacy and usability of paywall models.

\subsection{Cookie Paywalls Through Users’ Eyes: Objectives, Choice Architecture, and Design Expectations}\label{discussion:rq1}

\noindent
\textbf{Perception of objectives.} Most participants viewed cookie paywalls as primarily driven by economic interests (\textbf{T1.1}), with websites as the clear financial beneficiaries. Although some mentioned regulatory compliance (\textbf{T1.3}) or privacy protection (\textbf{T1.2}), these reasons were far less cited. This imbalance suggests that users perceive the model as primarily profit-oriented rather than as a mechanism for ensuring legal compliance or safeguarding their data rights.



\noindent
\textbf{Perception of fairness.} Fairness perceptions were complex, fluid, and context-dependent. Among the 14 participants, 10 described cookie paywalls as unfair, while 8 found them fair. These numbers include overlaps, as several participants expressed both perspectives depending on the situation. The unfairness was largely tied to the absence of a “reject all” option (\textbf{T2.5}) and the sense of coercion. At the same time, some participants justified fairness by pointing to the user’s ability to leave the site (\textbf{T2.1}).

These contrasting perspectives reflect different interpretive frames: while participants viewed the binary “pay-or-ok” model as unfair, some considered the broader option of leaving the website as a form of fairness. This indicates that cookie paywalls are seen as fair only in a narrow sense, when opting out entirely remains a viable path. However, the prevalent feeling of being forced to choose, or having “no real choice” but to leave, points to a deeper issue: a potential failure to meet the GDPR’s requirement for consent to be freely given, as defined in Articles 4(11) and 7.

Moreover, our findings suggest a disconnect between perceived fairness and user behaviour: those who deemed cookie paywalls unfair often refused to pay, while those who considered them fair did not necessarily express willingness to pay likewise. This gap highlights that fairness judgments do not necessarily translate into acceptance or compliance, raising critical concerns about the legitimacy of consent mechanisms under such models.

\noindent \textbf{Expectations.} 
Participants shared a range of expectations about how cookie paywalls should function, spanning issues of transparency, control, and alternatives to the binary pay-or-ok model. While some expectations concerned immediate design choices, others reflected a broader desire for rethinking the underlying model to provide more meaningful options for users.

\textit{First}, participants wanted greater transparency and control over data practices (\textbf{T3.2}). They sought clarity on what data is collected/tracked, who accesses it, what benefits payment offers, and why selective cookie consent is not allowed, i.e. why there is a request to accept ``all''.


Our study supports the legal requirements of transparency of personal data practices because users also want such information. Under Article 5(1)(a) GDPR, personal data processing must be transparent to users, requiring websites to inform them about the types of data processed, recipients (Art. 14), the associated scope and consequences~\cite{Transparency29WP}, and risks (Recital 39, GDPR). In practice, however, users often ignore such information, leading to a privacy paradox, where stated intentions and actual behaviours diverge~\cite{barth2017privacy}.

Strycharz et al.~\cite{STRYCHARZ2021106750} found that while legal information for consent can empower users, it may also reduce perceived risk, leading to a control paradox; increased control over data sharing can raise users’ willingness to share.

Presentation and tone also mattered: participants reacted negatively to perceived manipulation in content and design (\textbf{T4.7}). While this reaction might have been influenced by our own design choices (see Section~\ref{results:RQ2}), it underscores a key insight: users valued transparency presented with clarity and honesty, not in ways that felt exaggerated or fear-inducing. This highlights that transparency alone is insufficient; how information is framed is equally critical for fostering trust and supporting informed, autonomous decisions.

A notable concern was the statement that “personal data may be shared with 300 vendors”. Though legally compliant, our participants perceived it as coercive, particularly when paired with a rigid `pay-or-ok' structure. Such framing led users to feel pressured to either pay or agree to extensive data sharing, undermining the notion of freely given consent.

These findings suggest that even when transparency requirements are met, users may still feel manipulated, pointing to the need for addressing not only formal compliance but also substantive fairness, and the real impact on user autonomy and rights~\cite{malgieri2020fairness}.

\textit{Second}, many participants (9 out of 14) strongly expected a \textbf{“Reject all”} option, which they saw as essential to fairness. They felt pressured by the binary structure of cookie paywalls and wanted the ability to avoid both paying and consenting (\textbf{T3.1}). They also called for more granular control over cookie preferences.

Courts and regulators have affirmed the legitimacy of placing a “reject all” button at the first layer of consent~\cite{austrian_dpa_dsb_cookie_2024}, which in cookie paywalls would mean rejecting all purposes. However, in practice, rejecting some purposes often triggered the reappearance of the paywall, suggesting users are not truly allowed to opt out. This reflects \textit{privacy theatre}~\cite{smart2022understanding}, where designs/technologies create the illusion of privacy protection while doing little or nothing to meaningfully safeguard user data. This illusion of choice conceals manipulative design and undermines freely given, meaningful consent. The frustration over limited control and inability to fully customise purposes also relates to GDPR’s requirements for granular consent (Recital 32) and purpose specification (Art. 5(1)(b))~\cite{Purpose-specification}.

\textit{Third}, participants expressed interest in \textbf{alternative models}, such as accessing content through \textit{contextualised ads} rather than consenting to tracking or paying (\textbf{T3.3}). Some saw contextual ads as a privacy-preserving compromise, aligning with the Norwegian Consumer Council’s report advocating tracking-free contextualised advertising as both GDPR-compliant and supportive of healthier business models~\cite{myrstad2021time}.

However, critics warn that contextual advertising may still enable profiling, especially when AI tools analyse content and user behaviour~\cite{Grafenstein-et-al-paywalls}. This enables neuroprogrammatic advertising, which targets individuals based on emotional states and mood profiles, raising concerns about the intrusiveness of these so-called privacy-friendly alternatives. Many vendors combine contextual and personal data~\cite{EUCom-awo}, blurring boundaries between targeting methods.

Relatedly, Kyi et al.~\cite{Lin-purposes} found that users often accept personalised ads not out of preference, but due to a perceived lack of alternatives. These findings question the legitimacy of expanding targeted advertising in any form and reflect a broader discomfort with the inflexible `pay-or-ok' structure, further discussed in our recommendations (Section~\ref{subsec:reco}).

Finally, a few participants  expressed that cookie paywalls should not exist at all (\textbf{T3.4}), viewing them as exploitative and fundamentally incompatible with how they believed the Internet should function.

\subsection{Drivers of pay‑or‑consent choices on cookie paywalls: factors shaping user decision‑making}\label{discussion:rq2}

\noindent \textbf{Pay decisions are driven by multiple factors in combination.} From the results of our study, there was no real `killer feature', and no unique factor that would sway participants towards paying, and eventually, most responses encompassed \textit{multiple factors} for a conditional answer: a cheap and easy way to subscribe to authentic content on a set of trustworthy websites would trigger a \textit{maybe I will pay}.

\noindent \textbf{Users resent cookie paywalls and would likely never pay.} 
Several participants across several focus groups repeated that they would not pay regardless of the situation (\textbf{T4.2}), suggesting a \textit{general sentiment of resentment towards the business model} of cookie paywalls (recall that economic reasons were most invoked by participants), in line with their perception of relative unfairness of cookie paywalls, and the expectation to be able to reject all cookies without detriment.
The fact that users consent to tracking to access content, since they cannot pay, could be framed as a sign of economic coercion, which fundamentally undermines the validity of consent under the GDPR.

\noindent \textbf{Access to exclusive content influences users' decisions to pay or to consent.}
The most impactful factor driving the behaviour of participants was the \textit{access to exclusive content} (\textbf{T4.1}), with several participants requesting guarantees behind the exclusiveness of this content.
Note that participants were as willing to pay as they were to accept consent to cookies to access content.
Several participants implied that, by expecting a ``reject all'' option, they would prefer not to have to make the choice between paying or consenting to access the desired content.
%
For those with sufficient time, looking for content elsewhere was also mentioned as an alternative, while paying was mostly an option for participants with sufficient financial means; accepting tracking being a sort of \textit{resignation} for those with neither time nor money.

\noindent \textbf{A cheap or fair price factor might influence a few users to pay.}
Some participants were willing to pay in case of a \textit{cheap or fair price} (\textbf{T4.3}). 
However, this factor was unsurprisingly articulated with their financial means, with less wealthy participants stating that they would not pay \textit{because they are poor}.
This economic barrier raises concerns about the scalability of this model: if users are reluctant to pay for one website at the moment of this study, they might not pay for several websites or other types of online services (such as social media) in the future if cookie paywalls become even more prevalent.

This factor should be interpreted in the light of another, albeit less cited factor on the possibility of accessing \textit{multiple services in a single pay-off} (potentially selected by the user) (\textbf{T4.17}).
Few participants declared that paying for a single website \textit{doesn't make sense} (P14, FG4), while very few suggested that a (cheap) subscription granting access to several websites (P2 (FG1) mentioned 50) could make the offer acceptable. This is currently a reality of several websites granting access up to 500 other websites under a single payoff subscription (see the case of the subscription management platform ``Contentpass'' used by several German-speaking websites)~\cite{content_pass_contentpass_2025}. 

In addition, the economic barrier mentioned prompts critical reflection on data protection and privacy as fundamental rights. When assuring these rights become something one must pay for, these models risk introducing or amplifying socioeconomic discrimination. Those with limited financial means are forced to consent and exchange their personal data to access content, while wealthier users can simply buy their way out of tracking. As such, the model divides users into two classes: those who can afford these rights and those who cannot. This has profound implications for the exercise of consent. Consent given under financial pressure, when users might have preferred to pay but could not, is not freely given in the legal 
sense. This undermines the very foundations of data protection requirements like freely given, informed consent, and the fairness principle, as enshrined in GDPR.  These findings challenge the fairness and legitimacy of monetising the right to data protection through paywalls, especially when such systems are deployed at scale.

\subsection{Recommendations}
\label{subsec:reco}

\noindent \textbf{Rethinking the business model.}
 Our study indicates that the `pay-or-ok' model is viewed as profit-driven, widely perceived as unfair, and not enabling users to make a meaningful choice. 
 Willingness to pay emerges only under low pricing and favourable economic conditions. 
 Even flexible models like multi-website bundled prices (e.g. ``one subscription unlocks 500 websites'')  do not assure a paying decision.
 These findings align with prior work confirming that users consent to tracking nearly 99\% of the time~\cite{muller2024paying}. 
In its current state, the model does not reflect users' expectations, nor the protection of their personal data. This is especially relevant given the increasing prevalence of this model~\cite{stenwreth2024or}. 
Regulators must critically reconsider the model's legitimacy in the light of users' expectations, 
adequately balancing the economic needs of the sectors concerned, the free circulation of information, and fundamental rights. At the minimum, essential services (e.g., political news, education, health) should be protected from these monetisation models that force people to choose between personal (and sensitive) data~\cite{wesselkamp2021tracking} and access.
Ultimately, these findings could support a reflection on how these models risk eroding trust in EU digital law, if users’ actual experiences with this model feel systematically rigged.
Considering future legislative efforts, we recommend that the future Digital Fairness Act (DFA)~\cite{euronews2025digitalfairness}  explicitly prohibits this model.


\noindent
\textbf{Third-option.} Several participants suggested an additional choice, such as having \textit{contextualised ads} only, or a \textit{reject all} button (\textbf{T3.3} and \textbf{T3.1}), as a possible middle ground.
%
These findings align with the recent EDPB opinion~\cite{edpb_opinion_2024}, which states that offering binary choices ``should not be the default way forward''.  
Instead, controllers should provide an equivalent alternative that neither requires a fee nor involves processing personal data for behavioural advertising.
Less intrusive advertising models, such as contextual, general, or topic-based advertising, are cited as viable examples.
Let us, however, note the following. 
As pointed out in Section~\ref{discussion:rq1}, contextualised advertising does not necessarily lead to a privacy-friendly business model; certain conceptions free of personal data collection (outlined by Forbrukerrådet~\cite{myrstad2021time}) may then provide a better alternative than others.
%
Also, the provision of contextualised ads must nonetheless respect the wish of users to 
be informed about why they see certain ads, be able to 
be provided with other ads, and their noninvasive integration within interfaces.

\noindent \textbf{Free Trial.} 
Participants proposed the integration of a \textit{free trial} that would make them more inclined to pay.
Access to a free trial and previews can be a means to assess whether the content is authentic and original (which is also a factor playing in favour of fairness \textbf{T2.4}).
This option is actually already available on some cookie paywalls, such as Contentpass (a leading SMP), but not on Freechoice (the other main SMP). 
%
The rise of AI-generated content reinforces the appeal of free trial models as an increasing number of websites now offer low-quality articles produced by Large Language Models (LLMs)~\cite{orland_over_2025}.   
Participants saw such content as emblematic of what they would not be willing to pay for, describing it as lacking in effort and human involvement. 
These concerns highlight a growing scepticism toward automated content and a corresponding expectation that paid content should reflect human labour, editorial oversight, and authenticity.

\noindent
\textbf{Criteria for exclusivity and authenticity.} Participants request mechanisms to ensure the \textit{exclusivity and authenticity of content}, which would make them perceive cookie paywalls as more fair (\textbf{T2.4}). Participants' need for reassurance from organisations regarding privacy-related decisions is a finding in line with previous research \cite{kulyk2023people}. User-facing accountability mechanisms that assure such unique attributes across online services are expected for these models. Hence,
regulators must define what constitutes exclusivity and authenticity of services. 
%
\textit{Exclusivity} should be understood as offering 
original reporting, proprietary material, and should not consist merely of aggregated third-party content, subject to 
independent audits or self-certifications. 
\textit{Authenticity}, in turn, requires that content be produced by identifiable and credible authors or creators, subject to editorial oversight or quality control. 
Where AI-assisted content generation is employed, clear labelling and editorial supervision must be ensured.

\noindent
\textbf{Subscription management.}
Trust in service providers heavily affects users' decisions, and users distrust payment mechanisms they perceive as unsafe (\textbf{T4.8}), although paying via credit card over encrypted communication -- a typical means to pay online -- is generally considered safe.
Offering familiar and trusted payment options 
can help boost user trust. 
In this line, it is relevant to minimise the amount of personal data required for the subscription and to ensure that users can easily cancel subscription renewals~\cite{subscriptionsDP}.
We further highlight the need to examine the role of SMPs, like Contentpass, especially when they bundle access to hundreds of sites and obscure the relationship between users and individual services.

\section{Conclusion}
This study explored user perceptions and decision-making around cookie paywalls, revealing context-dependent and often ambivalent views.
Participants larg\-ely perceived cookie paywalls as profit-driven, with fairness assessments shaped by factors, such as the availability of meaningful choices, transparency in data practices, and the authenticity of paid content.
While certain conditions like exclusive content, fair pricing, and trusted service providers could, in theory, motivate payment, most participants clearly expressed their reluctance or unwillingness to pay.

A strong expectation for alternatives, such as a “reject all” option or the ability to access content via contextual ads, emerged throughout the discussions. 
These findings raise important questions about the nature of meaningful consent and the broader implications of monetising the rights to data protection and privacy. 
When these rights hinge on financial means, there is a risk of reinforcing inequality, potentially challenging the freely given consent requirement under GDPR.
Future research could engage with service providers to better understand their motivations and incentives 
behind implementing cookie paywalls. Capturing the perspectives from both sides could help bridge the gap between user expectations and business realities. In addition, broader and more diverse participant samples, combined with quantitative methods, could help validate and expand on the insights of our study, revealing demographic patterns and further nuances in how users across different contexts perceive and respond to cookie paywalls.

\section{Acknowledgments}
This work was supported by the RENFORCE research group and was partially supported by the Wallenberg AI, Autonomous Systems and Software Program (WASP) funded by the Knut and Alice Wallenberg Foundation.
We are grateful for the great reflections of Itxaso Dominguez de Olazabal on this model.

\bibliographystyle{splncs04} 
\bibliography{biblio.bib}

\begin{thebibliography}{10}
\providecommand{\url}[1]{\texttt{#1}}
\providecommand{\urlprefix}{URL }
\providecommand{\doi}[1]{https://doi.org/#1}

\bibitem{EUCom-awo}
Armitage, C., Botton, N., Dejeu-Castang, L., Lemoine, L.E.C., Directorate-General~for Communications~Networks, C., Technology): Study on the impact of recent developments in digital advertising on privacy, publishers and advertisers – Final report. Publications Office of the European Union (2023). \doi{doi/10.2759/294673}

\bibitem{austrian_dpa_dsb_cookie_2024}
{Austrian DPA (DSB)}: A cookie banner’s first layer needs to contain a visually equivalent option to reject cookies, \url{https://gdprhub.eu/index.php?title=BVwG_-_W_108_2284491-1}

\bibitem{AW-24-popets}
Aziz, M.A.B., Wilson, C.: {Johnny Still Can’t Opt-out: Assessing the IAB CCPA Compliance Framework}. Proc. Priv. Enhancing Technol.  \textbf{2024}(4),  349--363 (2024). \doi{10.56553/popets-2024-0120}, \url{https://doi.org/10.56553/popets-2024-0120}

\bibitem{Bachelet-contractPayorOk}
Bachelet, V.: “pay-or-consent” and emerging trends in digital contract law. https://doi-org.utrechtuniversity.idm.oclc.org/10.54648/erpl2024047 (2024)

\bibitem{barth2017privacy}
Barth, S., De~Jong, M.D.: The privacy paradox--investigating discrepancies between expressed privacy concerns and actual online behavior--a systematic literature review. Telematics and Informatics  \textbf{34}(7),  1038--1058 (2017)

\bibitem{Biel-etal-24-JOLT}
{Bielova}, N., Santos, C., Gray, C.M.: Two worlds apart! {C}losing the gap between regulating {EU} consent and user studies. Harvard Journal of Law \& Technology (JOLT)  \textbf{37} (2024)

\bibitem{braun_2006_method}
Braun, V., Clarke, V.: Using thematic analysis in psychology. Qualitative Research in Psychology  \textbf{3}(2) (Jan 2006). \doi{10.1191/1478088706qp063oa}

\bibitem{content_pass_contentpass_2025}
{Content Pass}: contentpass, \url{https://www.contentpass.net/en/publications}

\bibitem{Meta-payorok}
D'Amico, A., Pelekis, D., Santos, C.T., Duivenvoorde, B.: Meta’s pay-or-okay model: An analysis under {EU} data protection, consumer and competition law  \textbf{2024},  254--272. \doi{10.71265/tkk29041}

\bibitem{ndss19}
Degeling, M., Utz, C., Lentzsch, C., Hosseini, H., Schaub, F., Holz, T.: {We Value Your Privacy... Now take some cookies: Measuring the GDPR's impact on web privacy}. In: Proceedings of the 26th Network and Distributed System Security Symposium (2019)

\bibitem{edpb_opinion_2024}
{EDPB}: Opinion 08/2024 on valid consent in the context of consent or pay models implemented by large online platforms, \url{https://www.edpb.europa.eu/system/files/2024-04/edpb_opinion_202408_consentorpay_en.pdf}

\bibitem{euronews2025digitalfairness}
{Euronews}: Eu digital fairness act is for business as much as consumers, says justice commissioner (2025), \url{https://tinyurl.com/4svy5das}, accessed: 2025-05-15

\bibitem{EUCOM-Meta}
{European Commision}: Commission finds apple and meta in breach of the digital markets act, \url{https://ec.europa.eu/commission/presscorner/detail/en/ip_25_1085}

\bibitem{Transparency29WP}
{European Commission}: {Guidelines on transparency under Regulation 2016/679, WP260 rev.01}. Available at \url{https://ec.europa.eu/newsroom/article29/item-detail.cfm?item_id=622227} (2018)

\bibitem{Purpose-specification}
Fouad, I., Santos, C., Al~Kassar, F., Bielova, N., Calzavara, S.: On compliance of cookie purposes with the purpose specification principle. In: 2020 IEEE European Symposium on Security and Privacy Workshops (EuroS\&PW). IEEE (2020)

\bibitem{factorInfluenceConsent-22}
Giese, J., Stabauer, M.: Factors that influence cookie acceptance: Characteristics of cookie notices that users perceive to affect their decisions. In: 9th International Conference on HCI in Business, Government and Organizations. p. 272–285 (2022)

\bibitem{dark_pattern_legal_req}
Gray, C.M., Santos, C., Bielova, N., Toth, M., Clifford, D.: Dark patterns and the legal requirements of consent banners: An interaction criticism perspective. In: 2021 CHI Conference on Human Factors in Computing Systems (2021)

\bibitem{imc20}
Hils, M., Woods, D.W., B\"{o}hme, R.: Measuring the emergence of consent management on the web. In: Proceedings of the ACM Internet Measurement Conference. p. 317–332. IMC '20 (2020)

\bibitem{acceptall}
Kampanos, G., Shahandashti, S.F.: Accept all: The landscape of cookie banners in greece and the uk. In: ICT Systems Security and Privacy Protection (2021)

\bibitem{Kollmann-pay-or-ok}
Kollmann, K.: Reconciling “pay or okay” models with the gdpr: The austrian dpa decision and other recent approaches in europe (2023). \doi{https://doi.org/10.21552/edpl/2023/2/15}

\bibitem{krueger_2002_fginterviews}
Krueger, R.: Designing and conducting focus group interviews  (2002)

\bibitem{KulykOksana-cookieperception}
Kulyk, O., Hilt, A., Gerber, N., Volkamer, M.: {``This website uses cookies"}: Users’ perceptions and reactions to the cookie disclaimer. In: EuroUSEC (2018). \doi{10.14722/eurousec.2018.23012}

\bibitem{kulyk2023people}
Kulyk, O., Renaud, K., Costica, S.: People want reassurance when making privacy-related decisions—not technicalities. Journal of Systems and Software  (2023)

\bibitem{Lin-purposes}
Kyi, L., Mhaidli, A., Santos, C.T., Roesner, F., Biega, A.J.: “it doesn’t tell me anything about how my data is used”: User perceptions of data collection purposes. In: Proceedings of the 2024 CHI Conference on Human Factors in Computing Systems. CHI '24, Association for Computing Machinery, New York, NY, USA (2024). \doi{10.1145/3613904.3642260}, \url{https://doi.org/10.1145/3613904.3642260}

\bibitem{lazar2017research}
Lazar, J., Feng, J.H., Hochheiser, H.: Research methods in human-computer interaction. Morgan Kaufmann (2017)

\bibitem{taming-cookie-monster}
Leenes, R., Kosta, E.: Taming the cookie monster with dutch law – a tale of regulatory failure. Computer Law \& Security Review  \textbf{31} (03 2015). \doi{10.1016/j.clsr.2015.01.004}

\bibitem{consentframing-22}
Ma, E., Birrell, E.: Prospective consent: The effect of framing on cookie consent decisions. In: Extended Abstracts of the 2022 CHI Conference on Human Factors in Computing Systems (2022)

\bibitem{popets2020}
Machuletz, D., Boehme, R.: {Multiple Purposes, Multiple Problems: A User Study of Consent Dialogs after GDPR}. In: Proceedings on Privacy Enhancing Technologies Symposium. vol.~2, pp. 481--498 (2020)

\bibitem{malgieri2020fairness}
Malgieri, G.: The concept of fairness in the gdpr: A linguistic and contextual interpretation. FAT* '20: Proceedings of the 2020 Conference on Fairness, Accountability, and Transparency  \textbf{36},  105374 (2020). \doi{10.1016/j.clsr.2019.105374}

\bibitem{do_cookie_banners_respect}
Matte, C., Bielova, N., Santos, C.: {Do Cookie Banners Respect my Choice? Measuring Legal Compliance of Banners from IAB Europe's Transparency and Consent Framework}. In: 2020 IEEE Symposium on Security and Privacy (SP) (2020)

\bibitem{Matt-etal-20-APF}
Matte, C., Santos, C., Bielova, N.: {Purposes in IAB Europe's TCF: which legal basis and how are they used by advertisers?} In: {APF 2020 - Annual Privacy Forum}. pp. 1--24. Lisbon, Portugal (Oct 2020), \url{https://inria.hal.science/hal-02566891}

\bibitem{Grafenstein-et-al-paywalls}
{Max von Grafenstein and Nina Elisabeth Herbort}: {Regulation of Online Advertising’ (Federation of German Consumer Organisations} (2024), \url{https://www.vzbv.de/sites/default/files/2025-02/vzbv-Gutachten_Expert-Opinion_Grafenstein_Herbort_Online-Advertising.pdf}

\bibitem{morel2023legitimate}
Morel, V., Santos, C., Fredholm, V., Thunberg, A.: Legitimate interest is the new consent-large-scale measurement and legal compliance of iab europe tcf paywalls. In: 22nd Workshop on Privacy in the Electronic Society (2023)

\bibitem{morel2022your}
Morel, V., Santos, C., Lintao, Y., Human, S.: Your consent is worth 75 euros a year-measurement and lawfulness of cookie paywalls. In: Proceedings of the 21st Workshop on Privacy in the Electronic Society. pp. 213--218 (2022)

\bibitem{muller2024paying}
M{\"u}ller-Tribbensee, T., Miller, K.M., Skiera, B.: Paying for privacy: Pay-or-tracking walls. arXiv preprint arXiv:2403.03610  (2024)

\bibitem{myrstad2021time}
Myrstad, F., Tj{\o}stheim, I.: Time to ban surveillance-based advertising. the case against commercial surveillance online  (2021)

\bibitem{dark_patterns_after_gdpr}
Nouwens, M., Liccardi, I., Veale, M., Karger, D.R., Kagal, L.: Dark patterns after the {GDPR:} scraping consent pop-ups and demonstrating their influence. CoRR  \textbf{abs/2001.02479} (2020), \url{http://arxiv.org/abs/2001.02479}

\bibitem{orland_over_2025}
Orland, K.: Over half of {LLM}-written news summaries have “significant issues”—{BBC} analysis, \url{https://tinyurl.com/yzxxc27z}

\bibitem{rasaii2023thou}
Rasaii, A., Gosain, D., Gasser, O.: Thou shalt not reject: analyzing accept-or-pay cookie banners on the web. In: Proceedings of the 2023 ACM on Internet Measurement Conference. pp. 154--161 (2023)

\bibitem{asiaccs19}
Sanchez-Rola, I., Dell'Amico, M., Kotzias, P., Balzarotti, D., Bilge, L., Vervier, P.A., Santos, I.: Can i opt out yet? gdpr and the global illusion of cookie control. In: Proceedings of the 2019 ACM Asia Conference on Computer and Communications Security. p. 340–351. AsiaCCS '19 (2019)

\bibitem{Sant_etal_20_TechReg}
Santos, C., {Bielova}, N., Matte, C.: Are cookie banners indeed compliant with the law? {D}eciphering {EU} legal requirements on consent and technical means to verify compliance of cookie banners. Technology and Regulation (TechReg) pp. 91--135 (2020), \url{https://doi.org/10.26116/techreg.2020.009}

\bibitem{subscriptionsDP}
Sheil, A., Acar, G., Schraffenberger, H., Gellert, R., Malone, D.: Staying at the roach motel: Cross-country analysis of manipulative subscription and cancellation flows. In: Proceedings of the 2024 CHI Conference on Human Factors in Computing Systems. CHI '24, Association for Computing Machinery, New York, NY, USA (2024). \doi{10.1145/3613904.3642881}

\bibitem{smart2022understanding}
Smart, M.A., Sood, D., Vaccaro, K.: Understanding risks of privacy theater with differential privacy. Proceedings of the ACM on Human-Computer Interaction  \textbf{6}(CSCW2),  1--24 (2022)

\bibitem{nordichi20}
Soe, T.H., Nordberg, O.E., Guribye, F., Slavkovik, M.: Circumvention by design - dark patterns in cookie consent for online news outlets. In: 11th Nordic Conference on Human-Computer Interaction. NordiCHI '20 (2020)

\bibitem{stenwreth2024or}
Stenwreth, A., T{\"a}ng, S., Morel, V.: To be or not to be (in the eu): Measurement of discrepancies presented in cookie paywalls. arXiv preprint arXiv:2410.06920  (2024)

\bibitem{STRYCHARZ2021106750}
Strycharz, J., Smit, E., Helberger, N., {van Noort}, G.: No to cookies: Empowering impact of technical and legal knowledge on rejecting tracking cookies. Computers in Human Behavior  \textbf{120},  106750 (2021). \doi{https://doi.org/10.1016/j.chb.2021.106750}

\bibitem{4years}
Trevisan, M., Traverso, S., Bassi, E., Mellia, M.: {4 Years of EU Cookie Law: Results and Lessons Learned}. In: Privacy Enhancing Technologies 2019 (2019)

\bibitem{utz_informed_2019}
Utz, C., Degeling, M., Fahl, S., Schaub, F., Holz, T.: ({Un})informed {Consent}: {Studying} {GDPR} {Consent} {Notices} in the {Field}. Proceedings of the 2019 ACM SIGSAC Conference on Computer and Communications Security  (2019). \doi{10.1145/3319535.3354212}

\bibitem{wesselkamp2021tracking}
Wesselkamp, V., Fouad, I., Santos, C., et~al.: In-depth technical and legal analysis of tracking on health related websites with ernie extension. In: Proceedings of the 20th Workshop on Privacy in the Electronic Society (WPES '21). ACM, Seoul, Korea (2021). \doi{10.1145/3463676.3485603}

\end{thebibliography}

\appendix

\section{Focus group study materials and analysis}\label{appendix:figures}

\subsection{Demographic questions}\label{demog}
Here we list the demographic questions our participants were requested to answer. Note that all the questions had a ``Prefer not to answer'' option and therefore were optional. 

1) What is your age? 18-24, 25-34, 35-44, 45-54, 55-64, +65

2) How do you identify yourself? Female, Male, Non-binary

3) What is your latest educational degree/level? Middle School, High School, BSc, MSc, PhD

4) What is your latest educational degree/level? - What is (was) your field of study (if applicable)?

5) How would you rate your confidence in using computers and navigating the internet? 1 = Not confident at all, 2 = Slightly confident, 3 = Moderately confident, 4 =  Confident, 5 = Very confident. 

6) How would you describe your ability to understand and manage cookie consent banners? 1 = I don’t understand cookie banners and find them difficult to manage, 2 = I have a basic understanding of cookie banners but often struggle to manage them, 3 = I understand cookie banners well enough but sometimes need clarification, 4 = I am confident in understanding and managing cookie banners with minimal difficulty, 5 = I fully understand cookie banners and have no issues managing them.

7) What is your current work situation?	Employed or self-employed, Looking for a job, Studying, Others
8) What is your monthly income after tax? Less than 15,000 SEK, Between 15,000 and 25,000 SEK, Between 25,000 and 35,000 SEK, Between 35,000 and 45,000 SEK, More than 45,000 SEK. 

\subsection{Thematic analysis}\label{theme}
First, two authors independently read the transcripts to familiarise themselves with the content and held an initial discussion on excerpts relevant to the research questions, specifically: (1) factors influencing participants' behaviour toward cookie banners and paywalls, and (2) their perceptions and expectations of these mechanisms. Next, both authors individually coded the first focus group based on their discussion, and later refined and reviewed their codes in a joint meeting until they reached a consensus, resulting in a more structured codebook. Using this refined codebook, they proceeded with the remaining focus groups, repeating the process for each until all were analysed and a final, refined codebook was established. Finally, the codes were systematically merged into overarching themes through a collaborative effort.

\subsection{Mockups presented in the focus groups}
\begin{figure}
\ContinuedFloat*
\centering
\begin{subfigure}[b]{\textwidth}
   \includegraphics[width=1\linewidth]{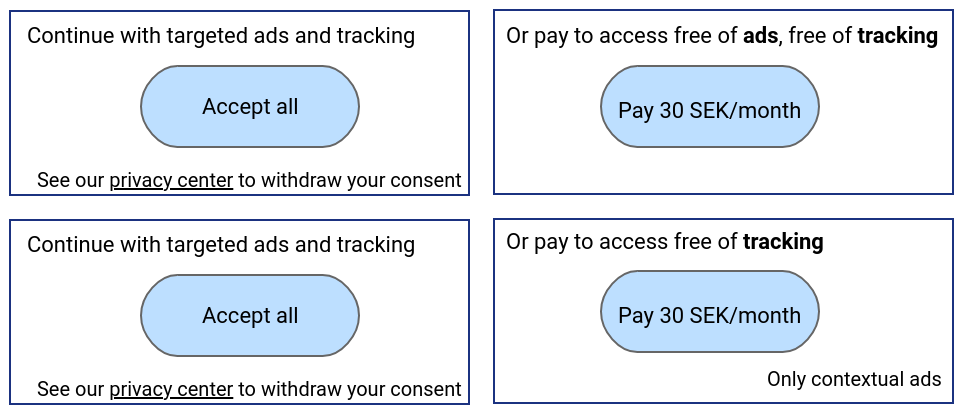}
   \caption{Scenario 1: two options (a) tracking-free and ad-free (top) and (b) ad-free only (bottom).}
   \label{fig:scenario_0} 
\end{subfigure}

\begin{subfigure}[b]{\textwidth}
   \includegraphics[width=1\linewidth]{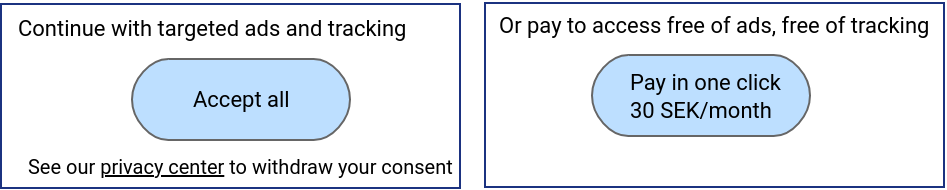}
   \caption{Scenario 2: One-click payment}
   \label{fig:scenario_1} 
\end{subfigure}

\begin{subfigure}[b]{\textwidth}
   \includegraphics[width=1\linewidth]{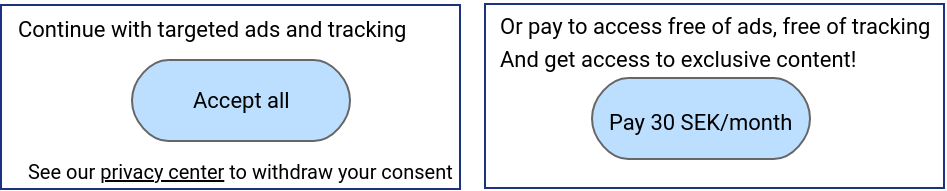}
   \caption{Scenario 3: Exclusive content}
   \label{fig:scenario_2}
\end{subfigure}

\caption{Mockups presented in our focus groups.}
\end{figure}

\begin{figure}
	\ContinuedFloat
\centering

\begin{subfigure}[b]{\textwidth}
   \includegraphics[width=1\linewidth]{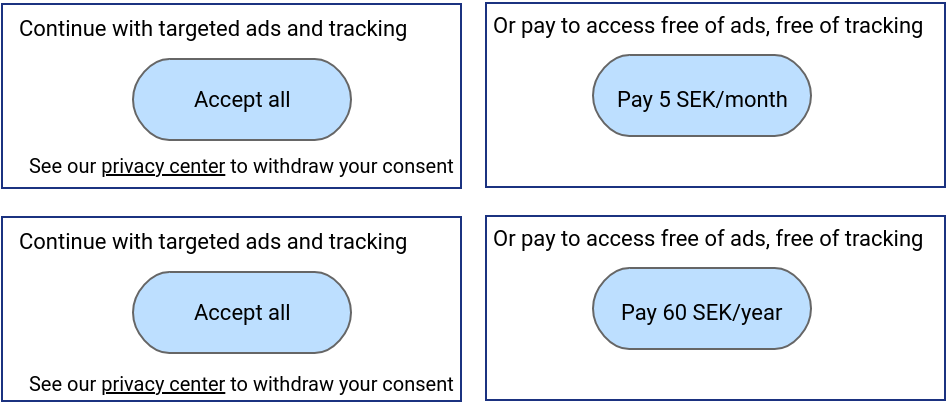}
   \caption{Scenario 4: Cheap price: a) monthly subscription, b) yearly subscription (no discount)}
   \label{fig:scenario_3}
\end{subfigure}

\begin{subfigure}[b]{\textwidth}
   \includegraphics[width=1\linewidth]{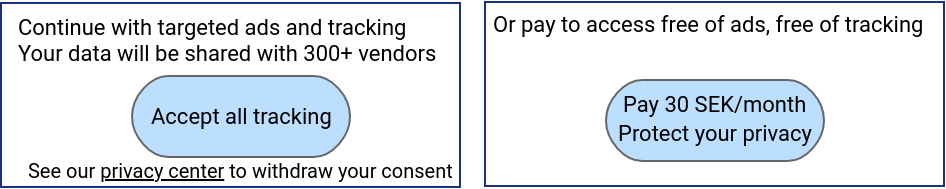}
   \caption{Scenario 5: Transparency on third-party information sharing}
   \label{fig:scenario_4}
\end{subfigure}

\begin{subfigure}[b]{\textwidth}
   \includegraphics[width=1\linewidth]{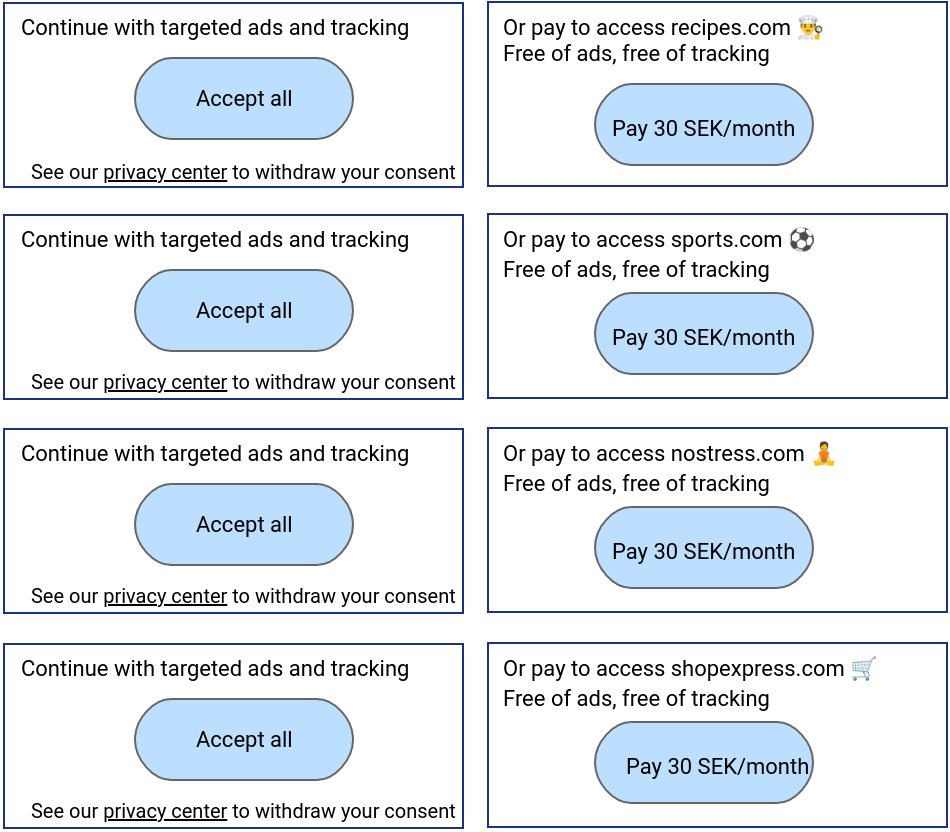}
   \caption{Scenario 6: Various types of websites}
   \label{fig:scenario_5}
\end{subfigure}

\caption{Mockups presented in our focus groups.}
\end{figure}

\begin{table}[]
\caption{Full codebook resulting from our thematic analysis}
\label{tab:codebook}
\hspace*{-1cm}
\begin{tabular}{@{}p{4cm}|l@{}}
\toprule
\textit{\textbf{Themes}} & \textbf{Codes} \\ \midrule
\multirow{3}{=}{\textbf{\textit{T1}} Users' perceptions of cookie paywalls' objective} & \textbf{T1.1} Monetary purpose for the websites - make money either way \\
 & \textbf{T1.2} To maintain users' privacy \\
 & \textbf{T1.3} To comply with regulations \\ \midrule
\multirow{6}{=}{\textbf{\textit{T2}} Perception of fairness of cookie paywalls} & \textbf{T2.1} Fair as users are not forced to take any options \\
 & \textbf{T2.2} Fair as nothing comes for free \\
 & \textbf{T2.3} Fair if transparent \\
 & \textbf{T2.4} Fair if authentic and original content \\ 
 & \textbf{T2.5} Unfair because forced to make a choice (can’t reject) \\
 & \textbf{T2.6} Unfair as privacy should not be a privilege \\ \midrule
 \multirow{4}{=}{\textbf{\textit{T3}} Expectations of cookie paywalls design} & \textbf{T3.1} A third option - Reject all \\
 & \textbf{T3.2} Better ex-ante transparency and control of data practices \\
 & \textbf{T3.3} Additional choice to have contextualized ads only \\
 & \textbf{T3.4} Cookie paywalls should not exist \\ \midrule
\multirow{21}{=}{\textbf{\textit{T4}} Factors affecting behaviour/decisions on cookie paywalls} & \textbf{T4.1} Access to exclusive content/service (regularly) \\
 & \textbf{T4.2} Not paying regardless of the situation \\
 & \textbf{T4.3} Cheap/fair subscription price \\
 & \textbf{T4.4} Not bothered/influenced by ads \\
 & \textbf{T4.5} Get rid of (blocking) ads \\
 & \textbf{T4.6} Feeling of pervasive tracking and data sharing \\
 & \textbf{T4.7} Perceived manipulation of content/design \\
 & \textbf{T4.8} Trust in the service provider \\
 & \textbf{T4.9} Feeling of discomfort because of wordings \\
 & \textbf{T4.10} Limited personal budget \\
 & \textbf{T4.11} Perceived unreasonable data collection for payment \\
 & \textbf{T4.12} Information should be free to access \\
 & \textbf{T4.13} Perceived feeling of safety because of other means in place \\
 & \textbf{T4.14} Lack of knowledge about cookie collection/tracking \\
 & \textbf{T4.15} Lack of perceived benefit from paying \\
 & \textbf{T4.16} Ex-ante transparency on data practices \\
 & \textbf{T4.17} Single pay-off, multiple services \\
 & \textbf{T4.18} Aversion towards subscriptions \\
 & \textbf{T4.19} Perceived sensitivity of context of service/content \\
 & \textbf{T4.20} Protect their privacy \\
 & \textbf{T4.21} Resentment of the business model \\ \bottomrule
\end{tabular}
\end{table}

\end{document}